\documentclass[12pt]{JHEP3} % pas draft
\usepackage{amsmath}
\usepackage{epsfig}
\input epsf.tex
\usepackage{amssymb,amsfonts}
\newcommand{\be}{\begin{equation}}
\newcommand{\ee}{{\end{equation}}}
\newcommand{\ba}{\begin{eqnarray}}
\newcommand{\ea}{{\end{eqnarray}}}

\renewcommand{\H}{{\cal H}}

\newcommand{\N}{{\cal N}}

\newcommand{\BZ}{\mathbb{Z}}

\newcommand{\Zop}{\mathbb{Z}}

\newcommand{\g}{\mathfrak{g}}

\newcommand{\rk}{{\rm rk}}

\newcommand{\su}{{\mathfrak{su}}}

\def\sqr#1#2{{
\vcenter{\vbox{\hrule height.#2pt
\hbox{\vrule width.#2pt height#1pt \kern#1pt
\vrule width.#2pt}
\hrule height.#2pt}}}}
\boldmath
\title{The coset D-branes of SU(n)}
\unboldmath
\author{Matthias R.\ Gaberdiel$^{1}$, Terry Gannon$^{2}$ and 
Daniel Roggenkamp$^{1}$
\\
$\ $ \\
$\ $ \\
$^1$Theoretische Physik, ETH-H\"onggerberg\\
\ CH-8093 Z\"urich, Switzerland\\
\ Email: \email{gaberdiel@itp.phys.ethz.ch}, 
\email{roggenka@phys.ethz.ch} \\
$\ $ \\
$^2$Department of Mathematical Sciences, University of
Alberta\\
\ Edmonton, Alberta, Canada, T6G 2G1\\
\ Email: \email{tgannon@math.ualberta.ca}
}

\abstract{Using a nested coset construction a collection of D-branes
that appear to generate all the K-theory charges of string theory on
${\rm SU}(n)$ are constructed and their charges are determined.}

\preprint{hep-th/0404112}

\begin{document}

\section{Introduction}

A lot of evidence has been accumulated over the last few years that
the charges of D-branes can be described 
in terms of (twisted) K-theory \cite{mm,WittenK,moore}. For example,
for the case of string theory on the simply connected Lie group $G$, the
charge group is conjectured to be the twisted K-group of $G$
(for more details see for example \cite{freed1,braun}). 
Modulo some technical assumption, this twisted K-group has been
calculated in \cite{braun} (see also \cite{freed2,fht}) to be  
\begin{equation}\label{chargegroup}
^{k+h^{\vee}}K^*(G)\cong \Zop_{d}^{\oplus m}\,,
\quad m=2^{\rk(\bar\g)-1}\,,
\end{equation}
where $h^\vee$ is the dual Coxeter number of $G$, $k$ is the level of
the underlying WZW model, and $d$ is the integer 
\begin{equation}
d = \frac{k+h^\vee}{{\rm gcd}(k+h^\vee,L)}\,.
\end{equation}
Here $L$ only depends on $\bar{\g}$, the finite dimensional Lie
algebra associated to $\g_k$. The summands $\Zop_{d}$ are 
equally divided between even and odd degree if $\rk(\bar\g)>1$. For 
$\bar\g=\mathfrak{su}(2)$ the only summand is in even degree. In this
paper we shall only consider the case of ${\rm SU}(n)$, for which 
$\rk(\su(n))=n-1$, and thus the multiplicity is $m=2^{n-2}$. 
\smallskip

On the other hand, D-branes can be constructed in terms of the
underlying conformal field theory, and it should be possible to
determine their charges using this microscopic description. In
particular, it was shown in \cite{fs} how the charges for branes
that preserve the affine algebra $\g_k$ (up to an automorphism) can in
principle be calculated. For the D-branes that preserve the full
algebra without any automorphism, the charges were then determined for
SU$(n)$ \cite{fs,mms}, and later for all  simply-connected Lie groups
\cite{bouwknegt}, and it was found that they account precisely for one
summand $\Zop_{d}$. The charges of the D-branes that preserve the
affine algebra up to an outer automorphism were calculated in
\cite{gg}; whenever these twisted D-branes exist, they also  
contribute one summand $\Zop_{d}$. These D-branes therefore only
account at most for two of the summands $\Zop_{d}$ in 
(\ref{chargegroup}).

Recently, a conformal field description of the remaining D-branes for
SU$(n)$ was proposed \cite{GGR}. This construction was inspired by the
suggestion of \cite{mms} that the remaining charges should be related
by some sort of T-duality to the original untwisted and twisted
branes. It was found that there are precisely $2^{n-2}$ 
different constructions that lead to boundary states that can be
distinguished by their coupling to the bosonic (and fermionic) degrees
of freedom associated to the Cartan torus. [In particular, these
D-branes therefore couple differently to the RR ground states of the
theory.] While these boundary states break in general the affine
symmetry, their open string spectra could still be described in terms
of twisted representations of the affine symmetry algebra. As a
consequence their charges could be determined, and it was found that
each of the $2^{n-2}$ different constructions leads to the same charge 
group $\Zop_d$.  

While this construction is quite suggestive, an obvious geometric
interpretation of the different D-branes was not available. 
In this paper we construct a different collection of D-branes whose
geometric interpretation is somewhat clearer. It is well known that
the group ${\rm SU}(n)$ is homotopy equivalent to a product of odd
dimensional spheres 
\begin{equation}
{\rm SU}(n)\cong S^{2n-1}\times S^{2n-3} \times \cdots \times S^3\,.
\end{equation}
Each of these spheres comes from a coset space
\begin{equation}
{\rm SU}(m) / {\rm SU}(m-1) \cong S^{2m-1}  \,,
\end{equation}
and thus, homotopically, we can think of ${\rm SU}(n)$ as the product 
\begin{eqnarray}\label{prspace}
{\rm SU}(n) & \cong &\Bigl({\rm SU}(n) / {\rm SU}(n-1) \Bigr) \times 
\Bigl({\rm SU}(n-1) / {\rm SU}(n-2) \Bigr) \times \\
& & \qquad\qquad\qquad\qquad\qquad\qquad\qquad \times \cdots \times
\Bigl({\rm SU}(3) / {\rm SU}(2) \Bigr) \times {\rm SU}(2)
\,. \nonumber 
\end{eqnarray}
In terms of the conformal field description of strings on 
${\rm SU}(n)$ this suggests that we should decompose the space of
states with respect to the corresponding coset algebras
(see also \cite{qs,quella}), {\it i.e.}
with respect to the $W$-algebra\footnote{Because of the
non-trivial $B$-flux, D-branes cannot wrap the sphere $S^3$; as
regards the construction of D-branes we therefore do not need to 
break the symmetry of the last ${\rm SU}(3)$ factor.}
\begin{equation}\label{walg}
{\cal W}=\widehat\su(n)/\widehat\su(n-1) 
\oplus \widehat\su(n-1)/\widehat\su(n-2)
\oplus\cdots\oplus \widehat\su(4)/\widehat\su(3)\oplus
\widehat\su(3)\,.
\end{equation}
The D-branes we construct respect this algebra, and they are therefore 
characterised by the gluing conditions that we impose for the
different factors. In fact, there are two different gluing conditions
that can be chosen for each factor, and this leads to $2^{n-2}$
different constructions. As we shall see, the construction of these
boundary states (and in particular their open string spectrum)
decouples for the various factors in ${\cal W}$. 
It is then suggestive 
to believe that the two constructions for each factor correspond
essentially to the choice of  whether or not the 
the homology class of the respective sphere contributes to the homology
class of the D-brane world volume.\footnote{While we are not able to
make this precise at present, we can show, following the analysis of
\cite{quella}, that at 
least some of these constructions correspond to branes wrapping
homologically inequivalent cycles; this will be discussed in
section~3.} Given the relation between K-theory and homology
\cite{mms} (namely that the summands of (\ref{chargegroup}) are in one
to one correspondence with the $2^{n-2}$ different homology cycles
that can be wrapped by D-branes), this then suggests that the above
construction accounts for all K-theory charges. 

Intriguingly, this construction is technically in many respects very
similar indeed to the construction of \cite{GGR}. This gives support
to the assertion that the branes described here represent different
points on the same sheets of moduli space as the branes of \cite{GGR}.
\smallskip

The paper is organised as follows. In the following section we
describe the construction of the D-branes, generalising the
construction of \cite{qs,quella}. We show that their open 
string spectrum gives rise to a NIM-rep of the product of coset
algebras. Furthermore, we identify flows that reduce the charges that
are carried by these D-branes to summands $\Zop_d$. In section~3 we 
show, following the analysis of \cite{quella}, that at least some of
these branes wrap homologically inequivalent cycles. Section~4
contains some conclusions.

\section{The inductive coset argument}
\setcounter{equation}{0}

String theory on a group manifold $G$ is described in terms of
representations of the affine Lie algebra $\g$ at level $k$. For the
situation where the group manifold is simply-connected (a systematic
analysis of the D-brane charges on non simply-connected Lie groups was
recently started in \cite{ggone,bs,fred}), the full spectrum of the
theory is 
then
\begin{equation}\label{spectrum}
\H = \bigoplus_{\lambda\in P^+(\g)} \H_\lambda \otimes
\bar\H_{\lambda^\ast} \,,
\end{equation}
where the sum runs over all integrable highest weight representations
$\lambda$ of $\g_k$, and the representation for the right-movers is
conjugate to the representation of the left-movers. (This theory is
therefore sometimes referred to as the `charge-conjugation' theory.)
In this paper we shall deal with the case when $\bar\g=\su(n)$. 

For $\bar\g=\su(2)$, the usual untwisted D-branes, {\it i.e.} the
D-branes that preserve the affine symmetry without any
automorphism\footnote{D-branes that preserve the affine symmetry up to
an inner automorphism can be related to one another by the group
action. All of these D-branes therefore carry the same charge; in the
following we can thus ignore such inner automorphisms.},
generate already the full charge group $\BZ_d$. The corresponding
boundary states are labelled by  the integrable highest weights
$\rho\in P^+_2:= P^+(\su(2)_k)$ 
(here as in the following the index `2' refers to the fact that we are
dealing with $\su(2)$, and we are suppressing the dependence on $k$),
and they are  explicitly given by the familiar Cardy formula
\begin{equation}
|\!| \rho_2 \rangle\!\rangle = \sum_{\lambda_2\in P^+_2} 
\frac{S_{\rho_2\lambda_2}}{\sqrt{S_{0\lambda_2}}} \, 
|\lambda_2\rangle\!\rangle^{\rm id}\,,
\end{equation}
where $S_{\rho\lambda}$ is the modular $S$-matrix, and  
$|\lambda\rangle\!\rangle^{\rm id}$ denotes the canonically normalised
Ishibashi state with trivial gluing conditions 
in the $\H_\lambda\otimes\bar\H_{\lambda^\ast}$ 
sector. The overlap between two such branes leads to open string
multiplicities that are simply described by the fusion rules,
\begin{equation}
\langle\!\langle \rho_2 |\!| q^{L_0+\bar{L}_0 - \frac{c}{12}} 
|\!| \rho_2' \rangle\!\rangle = 
\sum_{\lambda_2 \in P^+_2} N_{\rho_2 \lambda_2}{}^{\rho_2'} \,
\tilde\chi_{\lambda_2} \,,
\end{equation}
where $\tilde\chi_\lambda$ is the character of the $\lambda$
representation in the open string description. 

For $\bar\g=\su(3)$, the analogue of the above construction (that we
shall call the `straight' construction in the following) exists,
but there is also a second `twisted' construction, which  
accounts for the second $\BZ_d$-summand in the charge group 
$\BZ_d\oplus\BZ_d$. 
Namely, $\mathfrak{su}(3)$ has a non-trivial outer automorphism $\omega$
corresponding to complex conjugation, which can be used to twist the
gluing conditions. Boundary conditions obeying $\omega$-twisted gluing
conditions are labelled by the twisted representations $x_3$ of
$\widehat\su(3)$, and the corresponding boundary states are explicitly
given as 
\begin{equation}
|\!| x_3 \rangle\!\rangle = \sum_{\lambda_3\in P^-_3} 
\frac{\psi_{x_3\lambda_3}}{\sqrt{S_{0\lambda_3}}} \, 
|\lambda_3\rangle\!\rangle^\omega\,,
\end{equation}
where the $\psi$-matrix can be identified with a twisted $S$-matrix 
(see for example \cite{ggzero}), and $\lambda_3\in P^-_3$ denotes the
charge conjugation invariant weights of $\widehat\su(3)_k$, which give
rise to $\omega$-twisted Ishibashi states
$|\lambda_3\rangle\!\rangle^\omega$. 
These D-branes lead to open string multiplicities of the form
\begin{equation}
\langle\!\langle x_3 |\!| q^{L_0+\bar{L}_0 - \frac{c}{12}} 
|\!| x_3' \rangle\!\rangle = 
\sum_{\lambda_3 \in P^+_3} \N_{x_3 \lambda_3}{}^{x_3'} \,
\tilde\chi_{\lambda_3} \,,
\end{equation}
where $\N_{x \lambda}{}^{x'}$ are the twisted fusion rules.

For ${\rm SU}(2)$ and ${\rm SU}(3)$ these constructions already produce
D-branes that generate the full K-theory groups. However,
for ${\rm SU}(n)$ with $n\geq 4$, the K-theory analysis suggests a
charge group consisting of $2^{n-2}$ $\BZ_d$-subgroups, whereas there
are only two outer automorphisms (${\rm id}$ and $\omega$) of the
affine algebra. As we have mentioned before, inner automorphisms do
not modify the charges, and thus the D-branes that preserve the affine
symmetry (possibly up to an automorphism) can only account for two
summands in the charge group. We want to explain here how to construct
symmetry breaking D-branes that carry the remaining charges.

The idea of the construction is recursive. For each of the two
different constructions for $\widehat\su(3)$, there are two possible
constructions for $\widehat\su(4)$, depending on the choice of the
gluing condition for the coset algebra
$\widehat\su(4)/\widehat\su(3)$. We call the construction 
`straight' if we choose the trivial gluing condition for the coset
algebra, and `twisted' if the gluing condition is the one that is
induced from the outer automorphism $\omega$. Thus we get in total
four different constructions for ${\rm SU}(4)$. For each of these four
different constructions for ${\rm SU}(4)$ we have again two different
constructions for SU(5), depending on the choice of the gluing condition 
on the coset $\widehat\su(5)/\widehat\su(4)$, leading to a total of 
eight constructions
for ${\rm SU}(5)$, {\it etc}. Thus, for ${\rm SU}(n)$, we recursively
obtain $2^{n-2}$ different types of D-branes, which can precisely
account for the $2^{n-2}$ $\BZ_d$-summands in the charge group. 

More specifically, this construction leads to boundary conditions that
preserve the symmetry algebra
\begin{equation}\label{symmalg}
{\cal W}=\widehat\su(n)/\widehat\su(n-1) 
\oplus \widehat\su(n-1)/\widehat\su(n-2)
\oplus\cdots\oplus \widehat\su(4)/\widehat\su(3)\oplus
\widehat\su(3)\,,
\end{equation}
where the coset algebras $\widehat\su(l)/\widehat\su(l-1)$
come from the standard embeddings of $\widehat\su(l-1)$ into
$\widehat\su(l)$. [That is, the embedding is induced by the usual
embedding of $\su(l-1)$ into $\su(l)$, where the generators of
$\su(l-1)$ describe the upper-left block in $\su(l)$.]
They have the nice property that their selection
rules are trivial, {\it i.e.} every integrable highest 
weight representation  $\H_{\lambda_{l-1}}$ of $\widehat\su(l-1)$ 
appears in every integrable highest weight representation  
$\H_{\lambda_l}$ of $\widehat\su(l)$ 
\begin{equation}
\H_{\lambda_l}\cong
\bigoplus_{\lambda_{l-1}\in P^+_{l-1}}\H_{(\lambda_l,\lambda_{l-1})}
\otimes \H_{\lambda_{l-1}}\,,
\end{equation}
where the irreducible (non-trivial) representations of the coset
algebra are denoted by $\H_{(\lambda_l,\lambda_{l-1})}$.
Thus the representations of $\widehat\su(n)$ decompose as
\begin{equation}\label{decomp}
\H_{\lambda_n}\cong\bigoplus_{\lambda_i\in P^+_i,\;3\leq i<n}
\H_{(\lambda_n,\lambda_{n-1})}\otimes\H_{(\lambda_{n-1},\lambda_{n-2})}\otimes
\cdots\otimes\H_{(\lambda_4,\lambda_{3})}\otimes
\H_{\lambda_3}\,.
\end{equation}
We label such representations by 
$\hat\lambda=(\lambda_n,\lambda_{n-1},\ldots,\lambda_3)
\in P^+_n\times\cdots\times P^+_3=:{\cal P}$.

As explained above, we want to choose different combinations of gluing
conditions for ${\cal W}$. Each coset algebra
$\widehat\su(l)/\widehat\su(l-1)$
has a non-trivial automorphism, which is induced by the outer 
automorphism $\omega$ (complex conjugation) of $\widehat\su(l)$. 
(By abuse of notation,
we will denote it by the same symbol $\omega$.) This automorphism 
leaves the subalgebra $\widehat\su(l-1)$ invariant and actually
coincides with the outer automorphism of $\widehat\su(l-1)$ when
restricted to it. Thus, we will only use one symbol, $\omega$, for all
of the outer automorphisms. 

Taking into account that there are also two different gluing
conditions for $\widehat\su(3)$, there are two different gluing
conditions for each summand of ${\cal W}$. We can therefore describe
the entire gluing condition on ${\cal W}$ by 
$\hat\omega=(\omega_n,\ldots,\omega_3)$, where each $\omega_l$ is
either the trivial automorphism id, or the outer automorphism
$\omega$. For reasons that will become clear momentarily, it is
convenient to choose the convention that $\hat\omega$ describes the
gluing condition $\omega_n\,\omega_{n-1}\,\cdots\omega_l$ on 
$\widehat\su(l)/\widehat\su(l-1)$ (or on $\widehat\su(3)$ in the case
of $l=3$). [Here the product of the $\omega_m$ is simply the product
of the two commuting automorphisms id and $\omega$ with the relation
$\omega^2=$ id. So for example, for $\omega_n=\omega$, 
$\omega_{n-1}=$ id, we have $\omega_n\,\omega_{n-1}=\omega$, 
{\it  etc}.] 

To construct the corresponding boundary states we now have to
determine the possible Ishibashi states
$|\hat\lambda\rangle\!\rangle^{\hat\omega}$ 
for a given gluing condition $\hat\omega$. Because of (\ref{decomp})
one easily sees that there is exactly one such Ishibashi state for
all $\hat\lambda$ for which
\begin{eqnarray}
\omega_n(\lambda_n)=&\bar{\lambda}^*_n&=\lambda_n\,,\nonumber\\
\omega_n\cdots\omega_{l+1}(\lambda_l)=&\bar{\lambda}_l^*&=
\omega_{n}\cdots\omega_{l}(\lambda_l)\quad{\rm for}
\quad n>l\geq 3\,,
\end{eqnarray}
where ${\bar{\lambda}}_l$ denotes the corresponding representation of
the right movers, and $\ast$ denotes charge-conjugation. This
simplifies to (here we see the advantage of our
convention regarding the gluing condition) 
\begin{equation}
\omega_l(\lambda_l)=\lambda_l\quad{\rm for}
\quad n\geq l\geq 3\,,
\end{equation}
which are just the conditions for a representation of $\widehat\su(l)$
to be invariant under the twist $\omega_l$. It is convenient to
introduce signs $\epsilon_l$ that are defined to be $+1$ if
$\omega_l={\rm id}$ and $-1$ if $\omega_l=\omega$; the solution
of the above conditions are then exactly given by the 
$\hat\lambda\in P^{\epsilon_n}_n\times\cdots\times P^{\epsilon_3}_3
=:{\cal P}^{\hat\epsilon}$. Thus, for each 
$\hat\lambda\in{\cal P}^{\hat\epsilon}$  we obtain an Ishibashi state 
$|\hat\lambda\rangle\!\rangle^{\hat\omega}$ of gluing type
$\hat\omega$, which can be used to construct the boundary states. 

The construction of the boundary states can now be done as in 
\cite{qs,quella}.
This is to say, we construct boundary states for each $n-2$ tuple
$\hat\rho=(\rho_n,\ldots,\rho_3)$, where $\rho_l$ labels an untwisted
representation of $\widehat\su(l)$ if $\epsilon_l=+1$, and a 
twisted representation otherwise. They are explicitly given as 
\begin{equation}\label{bdstates}
|\!|\hat\rho\rangle\!\rangle^{\hat\omega}=
\sum_{\hat\lambda\in{\cal P}^{\hat\epsilon}}
\sqrt{S_{0\lambda_n}} \, 
\prod_{l=3}^n B_{\rho_l\lambda_l}
|\hat\lambda\rangle\!\rangle^{\hat\omega}\,,
\end{equation}
where the coefficients are
\begin{equation}
B_{\rho_l\lambda_l}=
\left\{\begin{array}{ll} 
{\displaystyle\frac{S_{\rho_l\lambda_l}}{S_{0\lambda_l}}}
\quad&{\rm if}\quad\epsilon_l=1\vspace{0.2cm} \\
{\displaystyle \frac{\psi_{\rho_l\lambda_l}}{S_{0\lambda_l}}}
& {\rm if}\quad\epsilon_l=-1\,,\end{array}\right. 
\end{equation}
and where $S$ and $\psi$ denote the untwisted and twisted
$S$-matrices of $\widehat\su(l)$, respectively (there should never be
any confusion as to which algebra they belong). Thinking of this 
construction recursively, this boundary state is obtained by using the
straight construction in going from $\widehat\su(l-1)$ to 
$\widehat\su(l)$ if $\epsilon_l=+1$, and the twisted construction
otherwise. 

The open string spectrum corresponding to these D-branes
consists of representations of ${\cal W}$, which are labelled
by the tuples 
\begin{equation}
(\nu,\sigma)=(\nu_n,\sigma_{n-1},\nu_{n-1},\sigma_{n-2},\ldots,
\nu_3,\sigma_3) \in P^+_{n}\times P^+_{n-1}\times P^+_{n-1}\times
\cdots\times P^+_3\times P^+_3\,.
\end{equation}
In particular, following the same arguments as
in \cite{qs,quella}, we obtain the open string spectrum  
between the D-branes corresponding to 
$|\!|\hat\rho\rangle\!\rangle^{\hat\omega}$
and $|\!|\hat\rho'\rangle\!\rangle^{\hat\omega}$
\begin{equation}
\N_{\hat\rho (\nu,\sigma)}
{}^{\hat\rho'}
= \N_{\rho_n \nu_n}^{\epsilon_n}{}^{\rho_n'} \,  
\prod_{l\ne n}  \left( \sum_{c_l\in P^+_l} 
N_{\nu_l {\sigma_l}^\ast}{}^{c_l}\, 
\N_{\rho_l c_l}^{\epsilon_l}{}^{\rho'_l} 
\right) \,,
\end{equation}
where we write $\N^-$ for the twisted fusion coefficients (the
NIM-rep) and $\N^+$ for the untwisted fusion coefficients.

In fact, it is not difficult to show that these numbers define a
NIM-rep of the symmetry algebra ${\cal W}$. Since the NIM-rep is a
product of factors that involve $\rho_l,\rho_l'$, it is
sufficient to show this separately for each of these factors. 
(In the following we drop the index $l$.) Then we compute
\begin{eqnarray}
\sum_{z\in P^\epsilon} \sum_{c\in P^+} N_{\nu\sigma^\ast}{}^{c}\,
\N_{x c}^\epsilon{}^{z} \,\sum_{d\in P^+} N_{\nu'\sigma'{}^\ast}{}^{d}
\,\N_{z d}^\epsilon{}^{y} & = & 
\sum_{e\in P^+}   \N_{x e}^\epsilon{}^{y} \, 
\sum_{c,d\in P^+} N_{\nu\sigma^\ast}{}^{c}\;
N_{\nu'\sigma'{}^\ast}{}^{d}\; N_{cd}{}^{e} \nonumber \\
& = & 
\sum_{e\in P^+}   \N_{x e}^\epsilon{}^{y} 
\sum_{c,d'\in P^+} N_{\nu\sigma^\ast}{}^{c}\;
N_{\nu'd'}{}^{e}\; N_{\sigma'{}^\ast c}{}^{d'} \nonumber \\
& = & 
\sum_{e\in P^+}   \N_{x e}^\epsilon{}^{y} 
\sum_{c',d'\in P^+} N_{c'\nu}{}^{d'}\;
N_{\sigma^\ast \sigma'{}^\ast}{}^{c'}\;
N_{d' \nu'}{}^{e} \nonumber \\
& = & 
\sum_{e\in P^+}   \N_{x e}^\epsilon{}^{y} 
\sum_{c',d''\in P^+} N_{c'{}^\ast d''}{}^{e}\;
N_{\sigma \sigma'}{}^{c'}\;
N_{\nu \nu'}{}^{d''}\,, \nonumber  
\end{eqnarray}
where we have used the NIM-rep property of each $\N^{\pm}$, 
\begin{equation}
\sum_{x'\in P^{\pm}} \N_{x\,b}^{\pm}{}^{x'}\, \N_{x'c}^{\pm}{}^{y} = 
\sum_{f\in P^+}
\N_{xf}^{\pm}{}^{y} \,N_{bc}{}^{f} \,,
\end{equation}
as well as $N_{\sigma^\ast \sigma'{}^\ast}{}^{c'}=
N_{\sigma \sigma'}{}^{c'{}^\ast}$. This establishes that we indeed get a
NIM-rep for the symmetry algebra ${\cal W}$, namely 
\begin{equation}
\sum_{\hat{z}\in {\cal P}^{\hat{\epsilon}}} \N_{\hat{x}(\nu,\sigma)}
{}^{\hat{z}}\, \N_{\hat{z}(\nu',\sigma')}{}^{\hat{y}} = \sum_{(d,c)}
N_{(\nu,\sigma)(\nu',\sigma')}{}^{(d,c)} \, 
\N_{\hat{x}(d,c)}{}^{\hat{y}} \,,
\end{equation}
where the fusion rules of the algebra ${\cal W}$ are given as 
\begin{equation}
N_{(\nu,\sigma)(\nu',\sigma')}{}^{(d,c)} = 
N_{\nu_n \nu'_n}{}^{d_n} \, 
\prod_{l\ne n}
\left( N_{\nu_l \nu'_l}{}^{d_l} \,N_{\sigma_l \sigma'_l}{}^{c_l}
\right) \,.
\end{equation}

In order to calculate the charge group carried by these branes, we
need to know the NIM-rep when restricted to the untwisted highest
weight representations of $\su(n)$, since we then get the identity
\begin{equation}
\dim(\lambda)\, q_{\hat\rho} = 
\sum_{\hat\rho'} \, 
\N_{\hat\rho (\lambda,{\bf 0})}
{}^{\hat\rho'} \, 
q_{\hat\rho'} \,,
\end{equation}
where $(\lambda,{\bf 0})$ is the representation of ${\cal W}$
for which $\nu_n=\lambda$, with all other representations equal to
zero. [This is the only natural way in which $\su(n)$ `sits inside'
${\cal W}$.] What is remarkable, and what allows one to determine
the charge group in this case, is that the NIM-rep simplifies for
these special weights to 
\begin{equation}\label{mess}
\N_{\hat\rho (\lambda,{\bf 0})}
{}^{\hat\rho'} = 
\N_{\rho_n \lambda}^{\epsilon_n}{}^{\rho_n'} \, 
\delta_{(\rho_{n-1},\ldots,\rho_3)}^{(\rho'_{n-1},\ldots,\rho'_3)} \,.
\end{equation}
It follows then by the usual argument
\cite{fs,mms,bouwknegt,gg} that the branes for a fixed 
$(\rho_{n-1},\ldots,\rho_3)$
give rise to a charge group of the correct size $\Zop_d$.  In
addition, there are presumably other flows that are not of this simple
type; they will give rise to relations among the branes differing in 
$\rho_{n-1},\ldots,\rho_3$. We conjecture that these flows will 
be sufficient to deduce that this multiplicity is actually trivial,
{\it i.e.} that all the branes with a given gluing condition
$\hat\omega$ which differ in all but the first labels actually carry
the same charge. This is plausible from the geometric point of view
described in the next section, where it is argued that the geometry of
the world-volume depends only on the gluing condition
$\hat\omega$. In particular, as is explained there, the different
boundary states where all signs $\epsilon_i=+1$, wrap the
homologically trivial class, and therefore the same class as the usual
untwisted (symmetric) D-branes of  ${\rm SU}(n)$. It is believed that
their charges are described by one summand of (\ref{chargegroup}), and
thus the D-branes that differ in $\rho_{n-1},\ldots,\rho_3$ cannot
generate new charges in this case. It then seems plausible that the
same will be true for all other collections of $\epsilon_i$ as well. 

In summary, these arguments suggest that we get one summand of
$\Zop_d$ for each combination of gluing conditions, thus producing
precisely the full charge group.

\section{Some comments about geometry}

Following the arguments in \cite{quella}, the world volumes of the
D-branes in ${\rm SU}(n)$ corresponding to the boundary states
(\ref{bdstates}) can be described by products of untwisted and twisted
conjugacy classes \cite{AS,fffs,sta}. To be more precise let us denote
by 
\begin{equation}
{\cal C}_l^+(h):=\{ghg^{-1}\,|\,g\in{\rm SU}(l)\}\,,\qquad
{\cal C}_l^-(h):=\{gh\bar{g}^{-1}\,|\,g\in{\rm SU}(l)\}
\end{equation}
the orbits of $h\in{\rm SU}(n)$ under the untwisted and twisted
adjoint actions of ${\rm SU}(l)$, which we take to be embedded into
${\rm SU}(n)$ in the same way as above. (Here $\bar{g}$ denotes the
complex conjugate of $g$.)
Then the world volume corresponding to
$|\!|\hat\rho\rangle\!\rangle^{\hat\omega}$ is given by
\begin{equation}
{\cal C}^{\epsilon_n}_n(h_{\rho_n})\cdots
{\cal C}^{\epsilon_3}_n(h_{\rho_3})\,,
\end{equation}
where $h_{\rho_l}=\exp\left(\frac{2\pi i}{k+l}
(\rho_l+\xi)\right)\in{\rm SU}(l)\subset{\rm SU}(n)$ with $\xi$ the
Weyl vector of ${\rm SU}(l)$ and $k$ the level of the underlying WZW
model.  

In principle we now have to determine the homology classes of 
${\rm SU}(n)$ represented by such world volumes. In general, this
seems to be quite difficult, and we will not attempt to do so
here. Instead, we will restrict to some simple special cases, and show     
that they give rise to linearly independent homology classes.  
 
In the geometric limit $k\rightarrow\infty$, all the points
$h_{\rho_l}$ converge to the identity $e$. In particular the untwisted 
conjugacy classes ${\cal C}^+_l(g_{\rho_l})$ therefore degenerate to
points, whereas the twisted conjugacy classes 
${\cal  C}^-_l(h_{\rho_l})$ converge to the images of the symmetric
spaces ${\rm SU}(l)/{\rm SO}(l)$ in ${\rm SU}(n)$ 
under the Cartan embedding $[g]\mapsto g\bar{g}^{-1}=gg^t$. 

Thus, the world volume corresponding to the boundary states with
trivial gluing condition $\hat\omega$, {\it i.e.} $\epsilon_i=1$ for
all $i$, is a point. To describe the homology classes represented by
D-branes with other gluing conditions, let us recall some facts about
the topology of ${\rm SU}(n)$. As was mentioned before, ${\rm SU}(n)$
is homotopy equivalent to a product of odd dimensional spheres 
\begin{equation}
{\rm SU}(n)\cong S^{2n-1}\times\cdots\times S^3\,,
\end{equation}
where each of the spheres comes from a homogeneous space 
$S^{2l-1}\cong {\rm SU}(l)/{\rm SU}(l-1)$. In particular the homology 
of ${\rm SU}(n)$ is given by the exterior algebra generated by the
sphere classes $w_{2l-1}$ of degree $2l-1$,
$H_*({\rm SU}(n);\BZ)\cong \Lambda[w_3,\ldots,w_{2n-1}]$.
(Recall that because of topological obstructions \cite{mms} classes
containing $w_3$ cannot be ``wrapped'' by D-branes but only those in  
$\Lambda[w_5,\ldots,w_{2n-1}]$.)

For odd $l$ (the analysis for even $l$ seems to be more complicated),
the homology of ${\rm SU}(l)/{\rm SO}(l)$ is given by 
$H_*({\rm SU}(l)/{\rm SO}(l);\BZ)\cong
\Lambda[e_5,e_9,\ldots,e_{2l-1}]$ (see {\it e.g.} III.6 of
\cite{math}). Furthermore, as can be deduced from the considerations
in \cite{pr}, the fundamental class 
$e_5\wedge e_9\wedge\cdots\wedge e_{2l-1}$ of ${\rm SU}(l)/{\rm SO}(l)$ 
is mapped under the Cartan embedding to a class proportional to
$w_5\wedge w_9\wedge\cdots\wedge w_{2l-1}$, which is thus the class
represented by ${\cal C}^-_l(e)\subset{\rm SU}(n)$. (A special case of 
this are the twisted D-branes of ${\rm SU}(3)$ whose
geometry has been discussed for example in \cite{mms}.) In particular,
it therefore follows that at least 
those constructions for which $\epsilon_i=1$ for all except for one
odd $i$ lead to
D-branes that wrap homologically inequivalent cycles. Since our
construction is in essence recursive, one would then expect that 
all classes in $\Lambda[w_5,\ldots,w_{2n-1}]$ can be represented 
by the world volumes of the various D-branes obtained by it.
Given the relation of K-theory to homology \cite{mms}, this would then
suggest that these boundary states indeed account for all the K-theory
charges.

\section{Conclusions}

In this paper we have given a construction of a collection of D-branes
for ${\rm SU}(n)$ that seems to account for all the K-theory charges
of ${\rm SU}(n)$. In particular, we have argued that the $2^{n-2}$ 
different constructions lead to D-branes that wrap homologically
different submanifolds of ${\rm SU}(n)$. Given the close relation
between homology and K-theory, it is then very plausible that the
corresponding D-branes carry inequivalent K-theory charges. We have
given evidence that each of the $2^{n-2}$ different constructions
leads to one summand of $\Zop_d$ in the K-theory charge group. Our
construction therefore seems to account for all K-theory charges.

The construction is technically very similar indeed to the
construction of the D-branes that was proposed in \cite{GGR}. In
particular, the different signs $s_m$ in \cite{GGR} correspond to
choosing the straight or twisted construction for
$\widehat\su(m)/\widehat\su(m-1)$, and so
$s_m=\epsilon_n\cdots\epsilon_m$. The first sign $s_n$ in that
construction determines whether the resulting NIM-rep  of
$\widehat\su(n)$ is the untwisted fusion rule or the twisted NIM-rep,
as in (\ref{mess}); its last sign $s_1$ is immaterial because of the
Weyl symmetry, which is also the reason why there is only one
construction for ${\rm SU}(2)$, and a reason we can stop (\ref{walg})
at $\widehat\su(3)$.  It is therefore very plausible that the 
branes of \cite{GGR} lie on the same sheets of moduli space as the
D-branes that have been constructed here. In the formulation of
\cite{GGR} the conformal field theory analysis of the charges was more
transparent (since the whole NIM-rep could be interpreted in terms
of twisted representations of $\widehat\su(n)$), whereas in the  
description given here the geometrical interpretation is more
suggestive. Taken together, the two constructions provide therefore
good evidence that either of them gives a construction of the D-branes
that carry all the K-theory charges of ${\rm SU}(n)$.

In this paper, we have only considered the case of ${\rm SU}(n)$, but
it should be straightforward (just as for the construction in
\cite{GGR}) to generalise it to other Lie groups. Given the structure
of our constructions, it seems likely that the corresponding D-branes
will again be numerous enough to account for all K-theory charges. It
would be interesting to check this in detail though.

\section*{Acknowledgments}

We thank Stefan Fredenhagen and Thomas Quella for useful
conversations. This work was begun while MRG and TG were visiting   
BIRS; we are very grateful for the wonderful working 
environment we experienced there! TG also thanks the hospitality of the
MPIM in Bonn, where he was when the paper was concluded.
The research of MRG is also
supported in part by the Swiss National Science Foundation, while that
of TG is supported in part by NSERC.


\begin{thebibliography}{99}

\bibitem{mm}
{\sc R.~Minasian, G.~Moore}, 
\emph{K-theory and Ramond-Ramond charge}, 
JHEP \textbf{9711} (1997) 002, 
{\tt hep-th/9710230}.

\bibitem{WittenK}
{\sc E.~Witten}, 
\emph{ D-branes and K-theory}, 
JHEP \textbf{9812} (1998) 019, {\tt hep-th/9810188}.

\bibitem{moore}
{\sc G.W.~Moore},
\emph{K-theory from a physical perspective},
{\tt hep-th/0304018}.

\bibitem{freed1}
{\sc D.S.~Freed},
\emph{The Verlinde algebra is twisted equivariant K-theory},
Turk.\ J.\ Math \textbf{25} (2001) 159, {\tt math.rt/0101038}.

\bibitem{braun}
{\sc V.~Braun},
\emph{Twisted K-theory of Lie groups},
{\tt hep-th/0305178}.

\bibitem{freed2}
{\sc D.S.~Freed},
\emph{Twisted K-theory and loop groups},
{\tt math.at/0206237}.

\bibitem{fht}
{\sc D.S.~Freed, M.J.~Hopkins, C.~Teleman},
\emph{Twisted K-theory and loop group representations},
{\tt math.at/0312155}.

\bibitem{fs}
{\sc S.~Fredenhagen, V.~Schomerus}, 
\emph{Branes on group manifolds, gluon condensates, and twisted
K-theory}, 
JHEP \textbf{0104} (2001) 007, {\tt hep-th/0012164}.

\bibitem{mms}
{\sc J.M.~Maldacena, G.W.~Moore, N.~Seiberg},
\emph{D-brane instantons and K-theory charges},
JHEP \textbf{0111} (2001) 062,
{\tt hep-th/0108100}.

\bibitem{bouwknegt}
{\sc P.~Bouwknegt, P.~Dawson, D.~Ridout},
\emph{D-branes on group manifolds and fusion rings},
JHEP \textbf{0212} (2002) 065,
{\tt hep-th/0210302}.

\bibitem{gg}
{\sc M.R.~Gaberdiel, T.~Gannon},
\emph{The charges of a twisted brane},
JHEP \textbf{0401} (2004) 018,
{\tt hep-th/0311242}.

\bibitem{GGR}
{\sc M.R.~Gaberdiel, T.~Gannon, D.~Roggenkamp},
\emph{The D-branes of SU(n)},
{\tt hep-th/0403271}.

\bibitem{qs}
{\sc T.~Quella, V.~Schomerus},
\emph{Symmetry breaking boundary states and defect lines},
JHEP \textbf{0206} (2002) 028, {\tt hep-th/0203161}.

\bibitem{quella}
{\sc T. Quella},
\emph{On the hierarchy of symmetry breaking D-branes in group
manifolds}, 
JHEP \textbf{0212} (2002) 009, {\tt hep-th/0209157}.

\bibitem{ggone}
{\sc M.R.~Gaberdiel, T.~Gannon},
\emph{D-brane charges on non-simply connected groups},
{\tt hep-th/0403011}, to appear in JHEP.

\bibitem{bs}
{\sc V.~Braun, S.~Sch\"afer-Nameki},
\emph{Supersymmetric WZW models and twisted K-theory of SO(3)},
{\tt hep-th/0403287}.

\bibitem{fred}
{\sc S.~Fredenhagen},
\emph{D-brane charges on SO(3)},
{\tt hep-th/0404017}.

\bibitem{ggzero}
{\sc M.R.~Gaberdiel, T.~Gannon},
\emph{Boundary states for WZW models}, 
Nucl.\ Phys.\ \textbf{B639} (2002) 471, 
{\tt hep-th/0202067}.

\bibitem{AS}
{\sc A.Yu.~Alekseev, V.~Schomerus},
\emph{D-branes in the WZW model},
Phys.\ Rev.\ \textbf{D60} (1999) 061901,
{\tt hep-th/9812193}.

\bibitem{fffs}
{\sc G.~Felder, J.~Fr\"ohlich, J.~Fuchs, C.~Schweigert},
\emph{The geometry of WZW branes},
J.\ Geom.\ Phys.\ \textbf{34} (2000) 162, {\tt hep-th/9909030}.

\bibitem{sta}
{\sc S.~Stanciu},
\emph{D-branes in group manifolds},
JHEP \textbf{0001} (2000) 025, {\tt hep-th/9909163}.


\bibitem{math}
{\sc M.~Mimura, H.~Toda},
\emph{Topology of Lie Groups. I, II},
Translations of Mathematical Monographs \textbf{91},
American Mathematical Society, Providence, RI (1991).

\bibitem{pr}
{\sc T.~P\"uttmann, A.~Rigas},
\emph{Presentations of the first homotopy groups of the unitary groups},
Comment. Math. Helv. \textbf{78} (2003) 648,
{\tt math.AT/0301192}.



\end{thebibliography}
\end{document}